\documentclass[twocolumn,a4paper,superscriptaddress,showpacs,floatfix,pra]{revtex4}

\usepackage{graphics}
\usepackage{amsmath}
\usepackage{graphicx}
\usepackage{amsfonts}
\usepackage{amssymb}
\usepackage{flafter}
\usepackage{afterpage}
\usepackage{color}

\DeclareSymbolFont{rsfs}{U}{rsfs}{m}{n}
\DeclareSymbolFontAlphabet{\mathcal}{rsfs}

\begin{document}

\title{Superconducting Analogues of Quantum Optical Phenomena: Macroscopic
Quantum Superpositions and Squeezing in a SQUID Ring.}
\author{M.~J.~Everitt}
\affiliation{Centre for Physical Electronics and Quantum Technology, School of Science and Technology, University of Sussex, 
Brighton, Sussex, BN1 9QT, U.K.}
\email{m.j.everitt@sussex.ac.uk}
\author{T.~D.~Clark}
\affiliation{Centre for Physical Electronics and Quantum Technology, School of Science and Technology, University of Sussex, 
Brighton, Sussex, BN1 9QT, U.K.}
\email{t.d.clark@sussex.ac.uk}
\author{P.~B.~Stiffell}
\affiliation{Centre for Physical Electronics and Quantum Technology, School of Science and Technology, University of Sussex, 
Brighton, Sussex, BN1 9QT, U.K.}
\author{A.~Vourdas}
\affiliation{Department of Computing, Bradford University, Bradford, West Yorkshire, BD7 1DP, UK.}
\author{J.~F.~Ralph}
\affiliation{Department of Electrical and Electronic Engineering, Liverpool University, Brownlow Hill, Liverpool, L69 3GJ, U.K.}
\author{R.~J.~Prance}
\affiliation{Centre for Physical Electronics and Quantum Technology, School of Science and Technology, University of Sussex, 
Brighton, Sussex, BN1 9QT, U.K.}
\author{H.~Prance}
\affiliation{Centre for Physical Electronics and Quantum Technology, School of Science and Technology, University of Sussex, 
Brighton, Sussex, BN1 9QT, U.K.}
\date{\today }

\begin{abstract}
In this paper  we explore the quantum behaviour of  a SQUID ring which
has a  significant Josephson  coupling energy. We  show that  that the
eigenfunctions of the  Hamiltonian for the ring can  be used to create
macroscopic  quantum superposition states  of the  ring. We  also show
that  the   ring  potential  may  be  utilised   to  squeeze  coherent
states. With  the SQUID  ring as  a strong contender  as a  device for
manipulating  quantum information,  such  properties may  be of  great
utility  in the  future. However,  as with  all candidate  systems for
quantum technologies,  decoherence is  a fundamental problem.  In this
paper  we  apply an  open  systems approach  to  model  the effect  of
coupling a  quantum mechanical  SQUID ring to  a thermal bath.  We use
this model to demonstrate the  manner in which decoherence affects the
quantum states of the ring.
\end{abstract}

\maketitle

\section*{Introduction}

In  two recent  publications~\cite{EverittSCVRPP01,EverittCSPPVR01} we
reported on the theoretical  description of a quantum mechanical SQUID
ring (here, a thick  superconducting ring enclosing a single Josephson
weak  link device)  coupled  to quantised  electromagnetic field  (em)
oscillator modes. In this work we emphasised that the SQUID ring could
be used  to control  various quantum phenomena  involving each  of the
circuit components of the coupled  system via the static magnetic bias
flux $\left(  \Phi _{x}\right) $  applied to the ring.  These included
frequency  conversion between  the em  modes and  quantum entanglement
extending across  the system, both with relevance  to emerging quantum
technologies              based              on              Josephson
devices~\cite{Chiorescu2003,Martinis2002,RouseHL95,SilvestriniRGE00,NakamuraCT97,NakamuraPT99,vanderWalWSHOLM00,lo_hk_1998,OrlandoMTvLLM99,MakhlinSS99,AverinNO90}. Furthermore,
work  by  Friedman  et  al  on SQUID  rings  has  highlighted  another
phenomenon  of  potentially  great  significance  to  these  incipient
technologies,   namely   the   creation   of   externally   controlled
superpositions of macroscopically distinct  states in a SQUID ring, or
other,       Josephson       weak       link      based,       circuit
configurations~\cite{FriedmanPCTL00}.  As  will  become apparent,  the
creation and  control of  such states is  a natural application  for a
SQUID ring.
\begin{figure*}[!t]
\begin{center}
\resizebox*{0.95\textwidth}{!}{\includegraphics{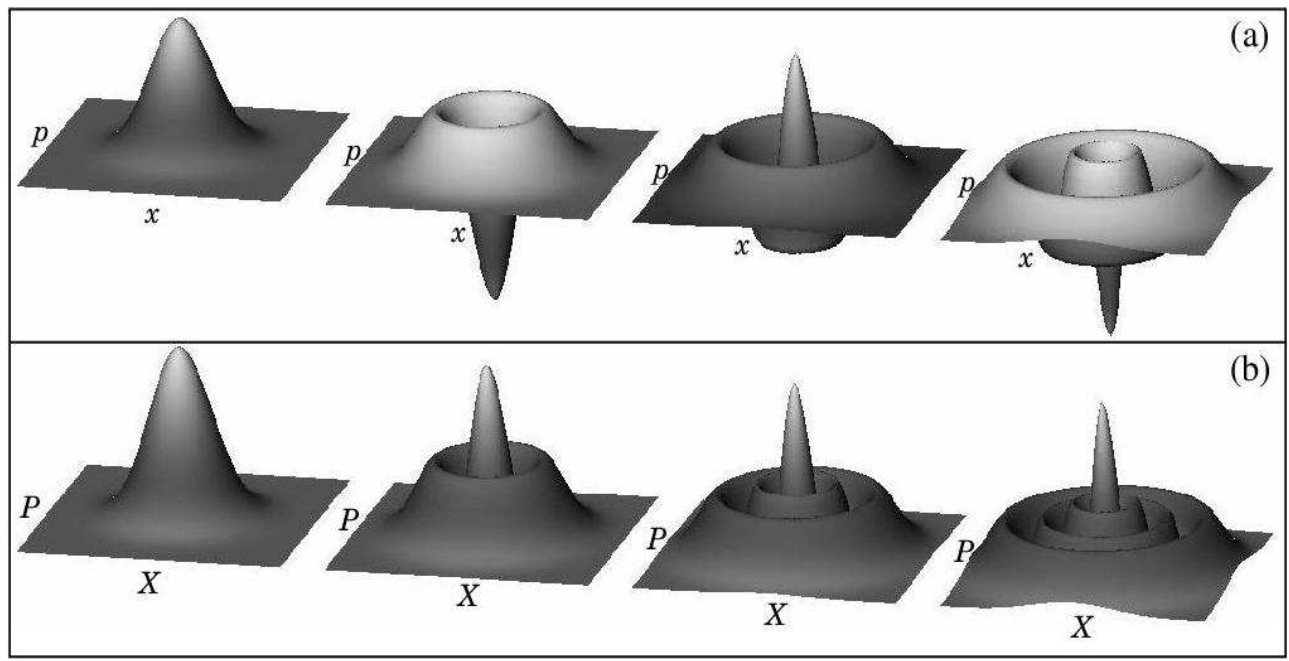}}
\end{center}
\caption{Illustrative example of (a)  Wigner and (b) absolute value of
the Weyl functions of the  first four energy eigenstates of the simple
harmonic oscillator (increasing in energy from left to right). }
\label{SHO}
\end{figure*}

In  this paper  we consider  the creation  and control  of macroscopic
quantum superposition states  in a SQUID ring alone,  uncoupled to any
em oscillator modes. First, we consider the spectral properties of the
ring Hamiltonian.  Then we observe  that at certain points in the bias
flux $\left( \Phi _{x}\right) $ applied to the ring the eigenfunctions
of  this Hamiltonian  form macroscopic  superposition states.  We show
that  a strong  enough level  of dissipation  may destroy  the quantum
nature of these states, whilst leaving the flux in the SQUID ring in a
statistical    mixture   of   two    macroscopically   distinguishable
states. Following this we demonstrate that a SQUID ring can be used to
create (form) a controllable  macroscopic superposition of states.  In
addition we show that a SQUID ring with a sufficiently large Josephson
coupling  term  in its  potential  can  be  used to  squeeze  coherent
states. In this it is apparent that physical phenomena associated with
SQUID rings, and with quantum  circuits built around SQUID rings, have
analogies   with  effects  well   known  in   the  field   of  quantum
optics. Indeed  the SQUID  ring can be  viewed as a  non-linear medium
which,  for  example,  can  be  utilised  to  generate  entanglements,
frequency conversion, superposition states and squeezing. However, the
SQUID  ring  has  significant  advantages over  the  generally  weakly
polynomial non-linear media of quantum optics which are usually weakly
coupled to external electromagnetic (em) fields. Thus, it is extremely
non-perturbative  in  nature (and  concomitantly  capable of  inducing
extremely                                                    non-linear
behaviour~\cite{EverittSCVRPP01,EverittCSPPVR01,Spi1992})    with    a
coupling to em modes that can be adjusted by means of an external bias
flux. This would appear to make the SQUID a prime candidate for future
developments in  what is,  in effect, highly  non-perturbative quantum
optics,  albeit   at  much   lower  frequencies.  In   practice  these
frequencies would typically  be much less than a  THz for low critical
temperature superconductors.

\section*{Background}

\subsection*{Wigner and Weyl Functions}

Although the Wigner  and Weyl functions are familiar  to those working
in             the             field            of             quantum
optics~\cite{Wigner1932,balazs1984,ChountasisV1998}, their  use in the
quantum description of Josephson weak link circuits, and in paqrticular
SQUID  rings, appears  to be  rather limited.  The Wigner  function is
defined to be
\begin{eqnarray*}
W\left( x,p\right) &=&\frac{1}{2\pi }\int d\zeta \left\langle x+\frac{1}{2}
\zeta \right\vert \rho \left\vert x-\frac{1}{2}\zeta \right\rangle \exp
\left( -i\zeta p\right) \\
&=&\frac{1}{2\pi }\int d\zeta \left\langle p+\frac{1}{2}\zeta \right\vert
\rho \left\vert p-\frac{1}{2}\zeta \right\rangle \exp \left( -i\zeta x\right)
\end{eqnarray*}
where  $\rho $ is  the density  operator describing  the state  of the
system  with  conjugate  variables  position $\left(  x\right)  $  and
momentum $\left(  p\right) $. Physically, the Wigner  function can, to
some extent, be considered as  a generalisation of the wavefunction of
the  quantum  system  under  study  in  which  we  are  provided  with
information  in both  position and  momentum space.  We note  that the
Wigner  function may,  and often  does, take  on negative  as  well as
positive values. An important and characteristic feature of the Wigner
function is that the  quantum correlations between the macroscopically
distinct components  of a macroscopic superposition state  can be seen
in an obvious  and graphical way, i.e. these  correlations will appear
in the Wigner function as interference terms between the states of the
superposition in the $x-p$ phase plane.

By contrast, the Weyl function is defined as 
\begin{eqnarray*}
\tilde{W}\left( X,P\right) &=&\frac{1}{2\pi }\int d\zeta \left\langle \zeta
+ \frac{1}{2}X\right\vert \rho \left\vert \zeta -\frac{1}{2}X\right\rangle
\exp \left( -i\zeta P\right) \\
&=&\frac{1}{2\pi }\int d\zeta \left\langle \zeta +\frac{1}{2}P\right\vert
\rho \left\vert \zeta -\frac{1}{2}P\right\rangle \exp \left( -i\zeta X\right)
\end{eqnarray*}
It is apparent here that the Weyl  function of a state is equal to the
overlap of the displaced state with the original state so that $X$ and
$P$ are considered  as increments in position and  momentum. As can be
seen, the Weyl function  is a generalised autocorrelation function; it
is  also   the  two  dimensional  Fourier  transform   of  the  Wigner
function. Thus, just as the  Wigner function highlights the regions in
the $x-p$  plane where the wavefunction amplitude  is significant, the
Weyl function tells  us where there exits a  significant amplitude for
correlations  between intervals  of  $\Delta x\left(  =X\right) $  and
$\Delta p\left( =P\right) $ in  this plane. A more detailed discussion
and review of Wigner and  Weyl functions, and the relationship between
them,          can          be          found          in          the
literature~\cite{ChountasisV1998,WolfgangS2001}.

For  those  unfamiliar with  these  functions  we  provide a  specific
example  in  figure~\ref{SHO}.  Here,   we  have  plotted  the  Wigner
functions for the first four energy eigenstates of the simple harmonic
oscillator together with the  absolute values of their associated Weyl
functions.  We note  that the  Weyl function  is, in  general, complex
valued. In this paper, therefore, we only ever plot its absolute value
since, for our purposes,  this provides us with sufficient information
about the correlations of the wavefunction.

\subsection*{The SQUID Ring Hamiltonian}

Over the last two decades  SQUID rings, viewed as single, macroscopic,
quantum  objects,  have been  the  subject  of considerable  attention
theoretically.  In  early studies  the  focus  was  primarily on  time
independent  properties  and  the  interaction  of  SQUID  rings  with
external  environments~\cite{Spi1992,SrivastavaW87,LeggettCDFGZ87}. Of
late there  has been  much interest in  time dependent  behaviour, for
example, in solving the time dependent Schr\"{o}dinger equation (TDSE)
for    a   SQUID    ring   in    the   presence    of    a   microwave
field~\cite{ClarkDREPPWWS98,DigginsWCPPRS97-b}.  Recently, significant
efforts have been devoted  to the experimental measurement and control
of    macroscopic    quantum    superposition    states    in    SQUID
rings~\cite{FriedmanPCTL00}.  In this  paper we  proceed to  develop a
theoretical description of macroscopic quantum superpositions in SQUID
rings,  borrowing on  techniques  that are  commonly  used in  quantum
optics. We  extend the usefulness  of this description  by considering
quantum  mechanical   squeezed  states   in  SQUID  rings.   For  both
superposition of states  and squeezing in these rings  we also discuss
the effect of dissipation (decoherence).

\begin{figure}[!t]
\begin{center}
\resizebox*{0.45\textwidth}{!}{\includegraphics{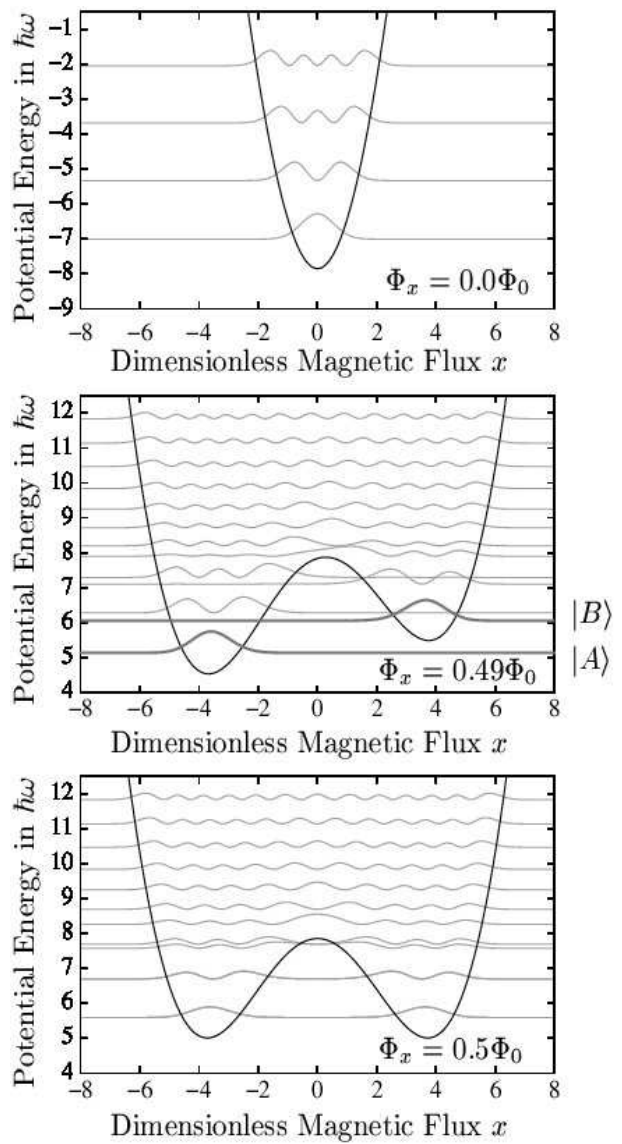}}
\end{center}
\caption{Potential energy in units  of $\hbar \protect\omega $ $\left(
\protect\omega    =1/\protect\sqrt{\Lambda    C}\right)    $    versus
dimensionless  flux (see  equation \protect\ref{posmom})  for  a SQUID
ring  with parameter values  $C=5\times 10^{-15}$F,  $\Lambda =3\times
10^{-10}$H  and $\hbar \protect\nu  =0.047\Phi _{0}^{2}/\Lambda  $ for
$\Phi _{x}=0,0.49$ and $0.5\Phi _{0}$.  Also shown are the probability
density functions of the rings wavefunctions displaced by their energy
eigenvalues.}
\label{f1}
\end{figure}
In   the   widely   used   lumped   component   model   of   a   SQUID
ring~\cite{Spi1992,SrivastavaW87} the Hamiltonian takes the form.
\begin{equation}
H=\frac{Q^{2}}{2C}+\frac{\left( \Phi  -\Phi _{x}\right) ^{2}}{2\Lambda
}-\hbar   \nu  \cos   \left(  \frac{2\pi   \Phi   }{\Phi  _{0}}\right)
\label{HamSQUID}
\end{equation}
where $\Phi $  and $Q$ are, respectively, the  magnetic flux threading
the ring and the electric displacement flux across the weak link (with
$\left[  \Phi ,Q\right]  =i\hbar  $),  $\hbar \nu  /2$  is the  matrix
element for Josephson tunnelling  through the weak link (with critical
current $I_{c}=2e\nu $), $\Phi  _{0}=h/2e$ is the superconducting flux
quantum and $\Lambda $ and  $C$ are, respectively, the ring inductance
and the capacitance of the weak link in the ring.

Introducing  a unitary  translation  operator $\mathbb{T}=\exp  \left(
-i\Phi  _{x}Q/\hbar  \right)  $,  we  can then  write  down  the  ring
Hamiltonian as
\begin{equation}
H^{\prime }={\mathbb{T}}^{\dagger }H{\mathbb{T}}=\frac{Q^{2}}{2C}+\frac{\Phi
^{2}}{2\Lambda }-\hbar \nu \cos \left( 2\pi \frac{\Phi +\Phi _{x}}{\Phi _{0}}\right)  \label{eq:HamST}
\end{equation}
where  it  is clear  that  as $\hbar  \nu  \rightarrow  0$ the  system
behaviour reduces to  that of a simple harmonic  oscillator. Given the
relation between our system and  the simple harmonic oscillator we now
define creation and annihilation operators in the usual way, i.e. as
\begin{equation*}
a=\sqrt{\frac{C\omega }{2\hbar }}\left( \Phi -\frac{i}{C\omega }Q\right)
,a^{\dagger }=\sqrt{\frac{C\omega }{2\hbar }}\left( \Phi +\frac{i}{C\omega }
Q\right) .
\end{equation*}
These raising and lowering operators,  as used in quantum optics, then
allow us to  write the ring Hamiltonian in a  more convenient form. We
also choose to express it in dimensionless units, normalised to $\hbar
\omega $, where  $\omega /2\pi =1/2\pi \sqrt{\Lambda C}$  is the SQUID
ring oscillator frequency. This takes the form
\begin{equation}
\mathcal{H}=\left( a^{\dagger }a+\frac{1}{2}\right) -\frac{\nu }{\omega }\cos \left( \frac{2\pi }{\Phi _{0}}\sqrt{\frac{\hbar }{2C\omega }}\left[
a+a^{\dagger }\right] +2\pi \varphi _{x}\right)  \label{HamNorm}
\end{equation}
where $\varphi _{x}=\Phi _{x}/\Phi _{0}$ is the normalised static bias
flux applied  to the SQUID ring. We  note that the cosine  term in the
Hamiltonian  introduces non-linearities  to all  orders. We  have seen
that   this   property   of   the   SQUID   ring   introduces   highly
non-perturbative  effects~\cite{EverittSCVRPP01,EverittCSPPVR01}  when
coupled to other  circuit systems. In this paper we  show that it also
gives  rise  to  quantum  superpositions of  macroscopically  distinct
states and squeezing within the ring itself.

From (\ref{HamNorm})  it is  apparent that as  long as the  ratio $\nu
:\omega $,  and the product $C\omega  \left( =\sqrt{C/\Lambda }\right)
$,  remain  the same  the  physics of  this  system  is unchanged.  We
therefore choose values of $C$  and $\hbar \upsilon $ $\left( \text{or
equivalently }I_{c}=2e\upsilon  \right) $  that can be  attained using
currently available micro-fabrication  techniques, that are physically
sensible and that  will lead to SQUID ring  systems exhibiting quantum
behaviour  at  experimentally   accessible  temperatures.  With  these
factors  in   mind  we   choose  the  circuit   parameters  $C=5\times
10^{-15}$\textrm{F},  $\Lambda   =3\times  10^{-10}$\textrm{H}  and  a
sufficiently   large   value   of   $\hbar   \nu   \left(   =0.047\Phi
_{0}^{2}/\Lambda ;  I_c = 2\mathrm{\mu  A}\right) $ to  generate clear
wells in  the ring potential. Thus,  for a thin  film Josephson tunnel
junction weak link with  a $1$nm oxide insulator thickness (dielectric
constant  $\approx $10)  a capacitance  of $5\times  10^{-15}$F yields
junction  dimensions  $\approx  0.25\mathrm{\mu  m}$  square,  readily
achieved  using micro-fabrication.  Again, with  these  dimensions the
supercurrent density  in the junction  is around $4\mathrm{kAcm}^{-2}$
which   is   perfectly   reasonable.  Furthermore,   with   $C=5\times
10^{-15}$\textrm{F},  $\Lambda =3\times 10^{-10}  $\textrm{H}, $\omega
/2\pi =130\mathrm{GHz}$, well below the frequency corresponding to the
superconducting    energy    gap    in   niobium    $\left(    \approx
1\mathrm{THz}\right)$,  a  metal  often   used  in  weak  link  device
fabrication. Given these chosen parameter values, and assuming, as our
example,  SQUID circuits based  on niobium,  these correspond  to $\nu
/\omega =7.9$  and $C\omega =4.1\times 10^{-3}$,  values which, unless
otherwise stated, we now keep to throughout this paper.

Adopting these values  of $\Lambda $, $C$ and $\hbar \nu  $ we show in
figure~\ref{f1}   the  SQUID  ring   potential  $U\left(   \Phi  ,\Phi
_{x}\right)  =\left( \Phi -\Phi  _{x}\right) ^{2}/2\Lambda  -\hbar \nu
\cos \left(  2\pi \Phi  /\Phi _{0}\right) $  - see  (\ref{HamSQUID}) -
computed  for   three  different   values  of  $\Phi   _{x}$  ($=0\Phi
_{0}\mathrm{\ (top)},0.49\Phi  _{0}\mathrm{\ (middle)} $  and $0.5\Phi
_{0}\mathrm{\ (bottom)}$). We also show in this figure the probability
densities  of the wavefunctions  of the  ring as  a function  of $\Phi
_{x}$.  These  probability densities  are  displaced  by their  energy
eigenvalues,  found by  solving the  time  independent Schr\"{o}dinger
equation (TISE) using the  Hamiltonian (\ref{HamSQUID}). As the bottom
of each of the local wells in the ring potential in figure~\ref{f1} is
approximately  quadratic, we  would expect  the solutions  deep within
these wells to look like  those for the simple harmonic oscillator. In
addition, we find that, on average, these states are slightly squeezed
in  terms of  the  magnetic flux  variable  $\Phi $.  For example,  in
figure~\ref{f1}  the lowest  state  in  each of  the  wells for  $\Phi
_{x}=0.49\Phi  _{0}$ has  $\left( \Delta  \left[  \sqrt{C\omega /\hbar
}\Phi  \right] \right)  ^{2}\approx  0.43$ and  $\left( \Delta  \left[
\sqrt{1/C\hbar \omega  }Q\right] \right) ^{2}\approx  0.58 $ (compared
to  $0.5$  in  the  unsqueezed  state),  where $\Phi  $  and  $Q$  are
normalised  to  $\left(  \hbar  /C\omega \right)  ^{\frac{1}{2}}$  and
$\left( \hbar C\omega \right) ^{\frac{1}{2}}$, respectively. To aid in
the description of the system  we adopt the term locally s-harmonic to
describe  this  behaviour,  where  the  prefix  \emph{s}  denotes  the
squeezed nature of these states.  The equivalence of the low lying set
of  energy  eigenvalues   deep  in  each  (local)  well   is  a  sound
approximation  except where  the energy  levels of  two or  more wells
align.  In   such  cases,  of  course,   symmetric  and  antisymmetric
superpositions of the eigenfunctions for the isolated wells develop.

\begin{figure}[!t]
\begin{center}
\resizebox*{0.45\textwidth}{!}{\includegraphics{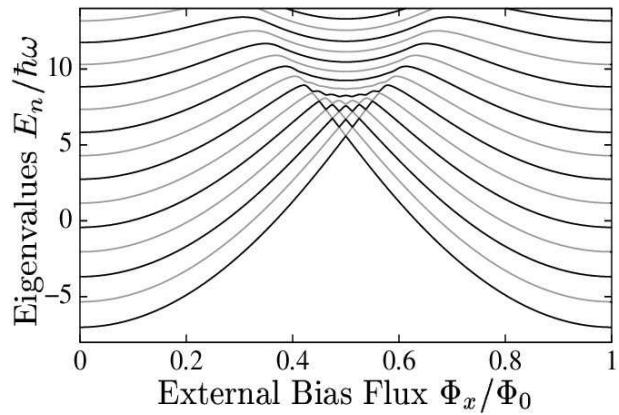}}
\end{center}
\caption{Computed SQUID ring energy eigenvalues of the Hamiltonian for the
SQUID ring of figure~\protect\ref{f1}. }
\label{f2}
\end{figure}

In figure~\ref{f2} 
we  show  a set  of  computed SQUID  ring  energy  eigenvalues of  the
Hamiltonian operator (\ref{HamNorm}),  starting with the ground state,
as  a function  of external  flux $\Phi  _{x}$ and  spanning  one flux
quantum. For clarity the energy  levels are shown as alternating black
and grey lines. As can be seen,  for the values of $\Lambda $, $C$ and
$\hbar \nu  $ we have adopted  in this work there  exist many crossing
points of  the original eigenvalues as  $\Phi _{x}$ is  changed over a
$\Phi  _{0}$ period. At  the scale  provided in  figure~\ref{f2} these
crossing points appear to be  degenerate in energy (e.g. in the lowest
two eigenvalues at  $\Phi _{x}=0.5\Phi _{0}$). Of course,  this is not
the case, as  would be evident if these  crossing points were computed
to  sufficient accuracy.  However, for  the potential  wells  shown in
figure~\ref{f1},  with  the  very  weak  coupling  between  levels  in
different  wells,  the  energy  splittings  at  these  points  may  be
extremely small indeed.

\section*{Results}

\subsection*{Quantum superpositions of macroscopically distinct states}

Ignoring,  for  the  moment,  the  special  cases  where  the  locally
s-harmonic  oscillator states  are degenerate  (zero  coupling between
wells) we instead consider making  a equal superposition of two energy
eigenstates. We  start by assuming  that these two states  are locally
s-harmonic in  different wells and, for simplicity,  take these states
to correspond to  the lowest energy levels in each  well. Then, as one
would expect,  we create  a superposition of  macroscopically distinct
states in flux. As this superposition state is no longer an eigenstate
of  the Hamiltonian for  the system,  it must  evolve with  time. This
evolution  introduces a  phase difference  between the  two stationary
(locally  s-harmonic)  states  in  the superposition  which  manifests
itself in a  time dependent evolution of the  interference term in the
Wigner function.

In computing Wigner  functions for the SQUID ring  it is convenient to
introduce  equivalent dimensionless  position $\left(  x\right)  $ and
momentum  $\left(  p\right)  $  operators  in place  of  $\Phi  $  and
$Q$. These are defined as
\begin{equation}
x=\sqrt{\frac{C\omega }{\hbar }}\Phi ,\,p=\sqrt{\frac{1}{\hbar C\omega }}Q
\label{posmom}
\end{equation}

\begin{figure}[!t]
\begin{center}
\resizebox*{0.45\textwidth}{!}{\includegraphics{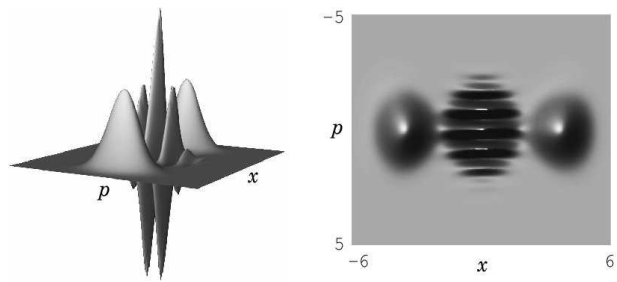}}
\end{center}
\caption{Wigner   function  showing,  both   in  perspective   and  in
projection  on  to  the  $x-p$  plane,  the  nature  of  this  quantum
superposition       of      macroscopically       distinct      states
$\frac{1}{\sqrt{2}}\left(    \left\vert   A\right\rangle   +\left\vert
B\right\rangle \right)  $ for the  SQUID ring of  figure~\ref{f1}. The
dimensionless    quantities    $x$   and    $p$    are   defined    in
equation~(\ref{posmom}) }
\label{f3}
\end{figure}

An example  of a macroscopic superposition  state in a  SQUID ring, as
illustrated  through  its  computed   Wigner  function,  is  shown  in
figure~\ref{f3},
both in  perspective and  in projection on  the $x-p$ plane.  Here, we
have taken  our standard values  of $\Lambda $,  $C$ and $\hbar  \nu $
(above)   and  have   selected  the   state  $\frac{1}{\sqrt{2}}\left(
\left\vert A\right\rangle +\left\vert  B\right\rangle \right) $ at the
flux  bias $\Phi  _{x}=0.49\Phi _{0}$  of figure~\ref{f1}(b).  In this
example we can distinguish  in the Wigner function two macroscopically
distinct flux  states of the SQUID  ring in the $x-p$  plane (i.e. the
gaussian-like   components),  separated   from  one   another   by  an
oscillatory region. The latter,  oscillatory, region arises because of
the  quantum coherence  between  the two  separate  components in  the
superposition and demonstrates that we  are indeed dealing with a true
quantum superposition  of macroscopically distinct states  in flux. We
note that  the Wigner function of figure~\ref{f3}  has been calculated
at a  fixed time $t=0$. However,  its general form does  not vary with
time.  Nevertheless, there is  dynamical evolution of the interference
term in the superposition but not in the observable flux states.

\subsection*{Tunable superposition of states in a SQUID ring}

\begin{figure}[!t]
\begin{center}
\resizebox*{0.45\textwidth}{!}{\includegraphics{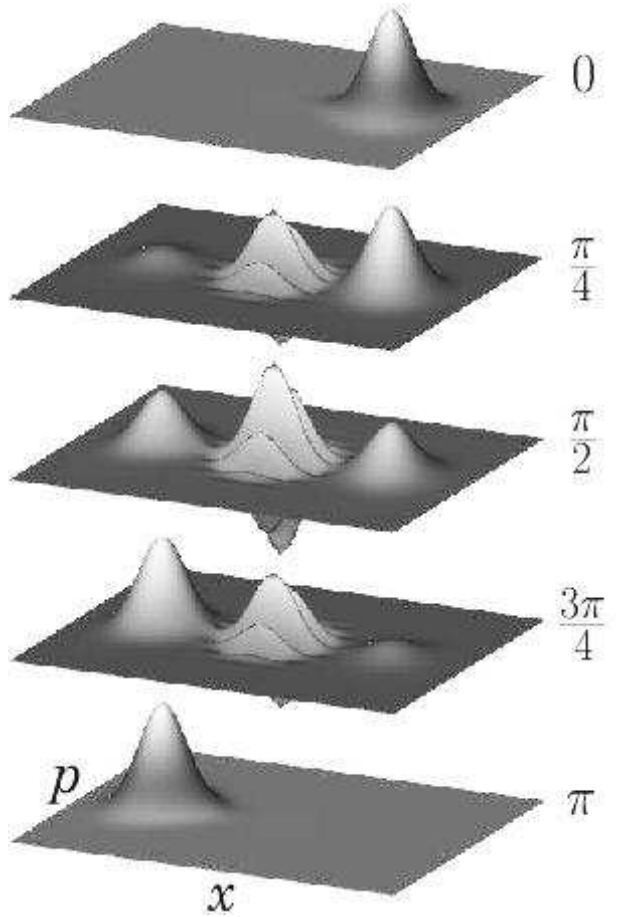}}
\end{center}
\caption{Wigner  function  of  a   superposition  of  the  lowest  two
(symmetric  and anti-symmetric)  energy eigenstates  as a  function of
phase  - see  equation~(\protect\ref{phaseequ}) and  related  text and
figure~\ref{f3} for scalings. }
\label{f4}
\end{figure}

We  now  consider  the  potential   with  a  static  bias  flux  $\Phi
_{x}=0.5\Phi   _{0}   $,   as    shown   in   the   bottom   plot   of
figure~\ref{f1}. By examining the first two energy eigenstates we find
that  we  have  wavefunctions   which  are  a  symmetric,  $\left\vert
s\right\rangle  $, and  antisymmetric,  $\left\vert a\right\rangle  $,
superposition  of the  lowest energy,  locally  s-harmonic, oscillator
vacuum states  of the  two middle wells.  These two states,  which are
squeezed,  are extremely  close in  energy due  to the  height  of the
barrier between the wells in  the potential. Even so, the interference
terms  in the Wigner  function between  the states  for each  of these
macroscopically distinct  states are $\pi  /2$ out of  phase. However,
from  the  viewpoint  of  theory  we  might  also  consider  an  equal
superposition of these states of the form
\begin{equation}
\frac{1}{\sqrt{2}}\left( \left\vert s\right\rangle +\exp (i\theta
)\left\vert a\right\rangle \right)  \label{phaseequ}
\end{equation}
where $\theta $ is the phase. Let $E_{s}$ and $E_{a}$ be the
eigenenergies  corresponding  to  the  eigenstates $|s\rangle  $  and
$|a\rangle  $ so  that  $\Delta E=E_{s}-E_{a}$  the energy  splitting
associated  with  the  lifting  of  the  degeneracy  concomitant  with
tunnelling between the adjacent wells. Then $\theta =\Delta Et$. It is
seen that this phase $\theta $  changes as a function of time and any
particular  value of $\exp  (i\theta )$  is realised  cyclically with
period $2\pi  /\Delta E$.  If we wished  to fix  our state in  a given
superposition we can do this  simply by changing our external magnetic
bias flux  by a fraction.  If this fractional  change in bias  flux is
sufficient this  will remove  the coupling between  the wells  and set
$\theta  $. The  effect  of changing  $\theta  $ can  be dramatic,  as
illustrated in figure~\ref{f4},
where the  Wigner function  in the $x-p$  plane is shown  for selected
values of  this parameter. As  is apparent, when the  wavefunction for
the ring  is strongly  (but not completely)  localised in two  or more
regions in  the ring potential  it is the quantum  interaction between
these  regions  that  is  responsible  for  the  creation  of superposition of macroscopically distinct states.

We note  that given the  very slightly different energies  between the
symmetric and antisymmetric superposition states, due to the height of
the barrier between  the two wells, the superposition~(\ref{phaseequ})
will oscillate slowly back and  forth between these wells. Using these
energies we  calculate this  period to be  100ns. This is  much longer
than the time  constant corresponding to the $\Lambda  C$ frequency of
the    SQUID   ring    in   our    example   (i.e.    for   $C=5\times
10^{-15}$\textrm{F},  $\Lambda =3\times  10^{-10}$ \textrm{H}  this is
$7.6$\textrm{ps})  but well  within  the decoherence  times of  modern
SQUID ring circuits~\cite{Martinis2002,Zhou2002,Han2001,Mooij2003}.

\begin{figure}[!t]
\begin{center}
\resizebox*{0.45\textwidth}{!}{\includegraphics{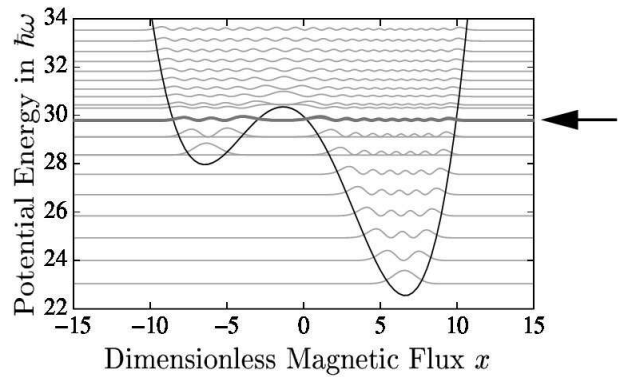}}
\end{center}
\caption{Potential   energy  in   units  of   $\hbar   \omega$  versus
dimensionless flux  (see equation~\ref{posmom})  of a SQUID  ring with
parameter  values  $  C=1.03\times  10^{-13}$F,  $\Lambda  =2.38\times
10^{-10}$H,  $I_{c}=2.02\times 10^{-6}$A  and  $\Phi _{x}=0.514466\Phi
_{0}$  (after  Friedmann  et  al~\protect\cite{FriedmanPCTL00}).  Also
shown are the probability density functions of the rings wavefunctions
displaced by their energy eigenvalues.  The arrow indicates the states
used to calculate the superposition state.}
\label{friedpot}
\end{figure}

In  the  above  discussion  we  have  considered  the  development  of
macroscopic superposition states in a  SQUID ring based on a choice of
ring parameters  which can  be realised by  fabrication and  which are
physically reasonable. However, this  choice does not connect directly
with published  experimental data.  We will therefore  deviate briefly
from our standard parameter values and consider an explicit example of
superpositions  of  SQUID ring  states  as  reported  recently in  the
literature by Friedman et  al \cite{FriedmanPCTL00}. In this paper the
authors  considered   a  \emph{``Quantum  Superposition   of  Distinct
Macroscopic States''}, and  presented experimental evidence indicating
that a SQUID ring could be placed into a superposition of two magnetic
flux states. We now demonstrate by computation that this superposition
may  indeed   form  a   true  macroscopic  superposition   state.  The
experimental  system used  a SQUID  ring with  the  circuit parameters
$C=1.03\times      10^{-13}$\textrm{F},      $\Lambda      =2.38\times
10^{-10}$\textrm{H}   and  $I_{c}=2.02\times   10^{-6}$\textrm{A}.  To
obtain a superposition state as demonstrated in this paper we used the
external bias flux quoted by the authors, i.e. $\Phi _{x}=0.514466\Phi
_{0}$. The potential  energy of a SQUID ring  with these parameters is
shown in figure~\ref{friedpot} together with the probability densities
for the ring  wavefunctions. These are displaced, as  before, by their
energy eigenvalues.  In this figure we  have marked with  an arrow the
states in the  left hand and right hand wells from  which we will form
our superpositions.  These are  the states that  were utilised  in the
Friedman et al experiment.

\begin{figure*}[!t]
\begin{center}
\resizebox*{0.95\textwidth}{!}{\includegraphics{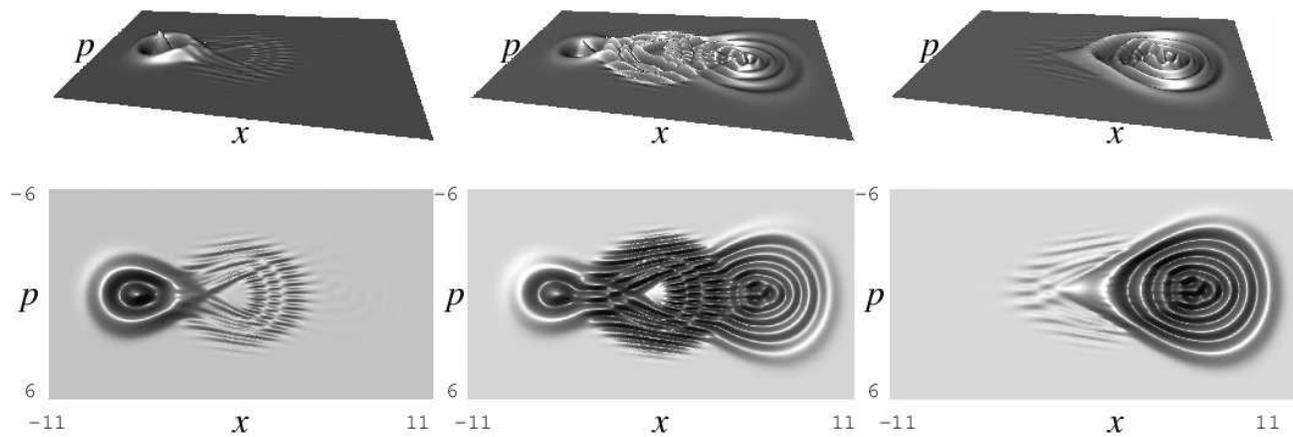}}
\end{center}
\caption{Wigner function,  from left to right, for  the three distinct
phases       $0,\protect\pi/2$       and      $\protect\pi$       (cf.
figure~\protect\ref{f4}       and       after       Friedmann       et
al~\protect\cite{FriedmanPCTL00}), showing, both in perspective and in
projection on to the $x-p$  plane, the superposition of the (symmetric
and    anti-symmetric)   energy    eigenstates    as   indicated    in
figure~\protect\ref{friedpot}.}
\label{friedwig}
\end{figure*}

These eigenfunctions are  similar to those of figure  \ref{f4} in that
they   form   a   symmetric,   $\left\vert   s\right\rangle   $,   and
antisymmetric, $\left\vert  a\right\rangle $, superposition  of the of
the  locally s-harmonic  states  of the  separate  wells, albeit  with
higher and different  ordinal numbers. The quantum state  of the SQUID
ring, as  reported by Friedman et  al.~\cite{FriedmanPCTL00} will thus
be a superposition  of these two eigenstates.  For  our purposes it is
therefore sufficient to look (again) at superpositions of the form
\begin{equation}
\frac{1}{\sqrt{2}}\left( \left\vert s\right\rangle +\exp (i\theta
)\left\vert a\right\rangle \right) .  \label{phaseequ2}
\end{equation}

In  figure~\ref{friedwig}  we  show  the Wigner  functions  for  three
different   superpositions  of  these   eigenstates,  again   both  in
perspective  and projection  on to  the $x-p$  plane.  We  notice that
compared with the Wigner  functions of figure~\ref{f4} both the states
of the macroscopic quantum superposition corresponding to the left and
right  hand   wells  in  figure~\ref{friedwig},  and   the  region  of
interference between  them, display more complex  patterns.  Given the
choice of more highly excited states in the Friedmann et al experiment
this   is  to  be   expected.   Nevertheless,   figure  \ref{friedwig}
demonstrates that the Wigner  function (and, of course, the associated
Weyl function) can expose  sophisticated quantum coherent behaviour in
SQUID rings.

\subsection*{Dissipation}

In considering the effect of dissipation on the calculations presented
in  this paper  we have  chosen to  use a  standard approach,  and one
familiar {in} quantum optics. This  is to done by coupling the {SQUID}
ring to  a decohering monochromatic thermal bath.  The master equation
for the evolution of the density operator of the system then takes the
form~\cite{weiss1999},
\begin{eqnarray}
\frac{\partial \rho }{\partial t}=-\frac{i}{\hbar }\left[ \mathcal{H},\rho \right] &+&\frac{\gamma }{2\hbar }\left( M+1\right) \left( 2a\rho a^{\dagger
}-a^{\dagger }a\rho -\rho a^{\dagger }a\right)  \notag \\
&&+\frac{\gamma }{2\hbar }M\left( 2a^{\dagger }\rho a-aa^{\dagger }\rho
-\rho aa^{\dagger }\right) .  \label{dis}
\end{eqnarray}
where $M$ is related to  the temperature $T$ and the frequency $\omega
_{b}$  of each  decohering bath  via $M_{i}=\left(  \exp  \left( \hbar
\omega _{b}/k_{B}T\right)  -1\right) ^{-1}$  and $\gamma _{i}$  is the
coupling  (damping  rate)  between  each  of  the  components  to  its
respective thermal bath. For the {following} examples that we will now
calculate we set $T=1\mathrm{K}$ with $\omega_{b}=\omega$.

\begin{figure}[!t]
\begin{center}
\resizebox*{0.45\textwidth}{!}{\includegraphics{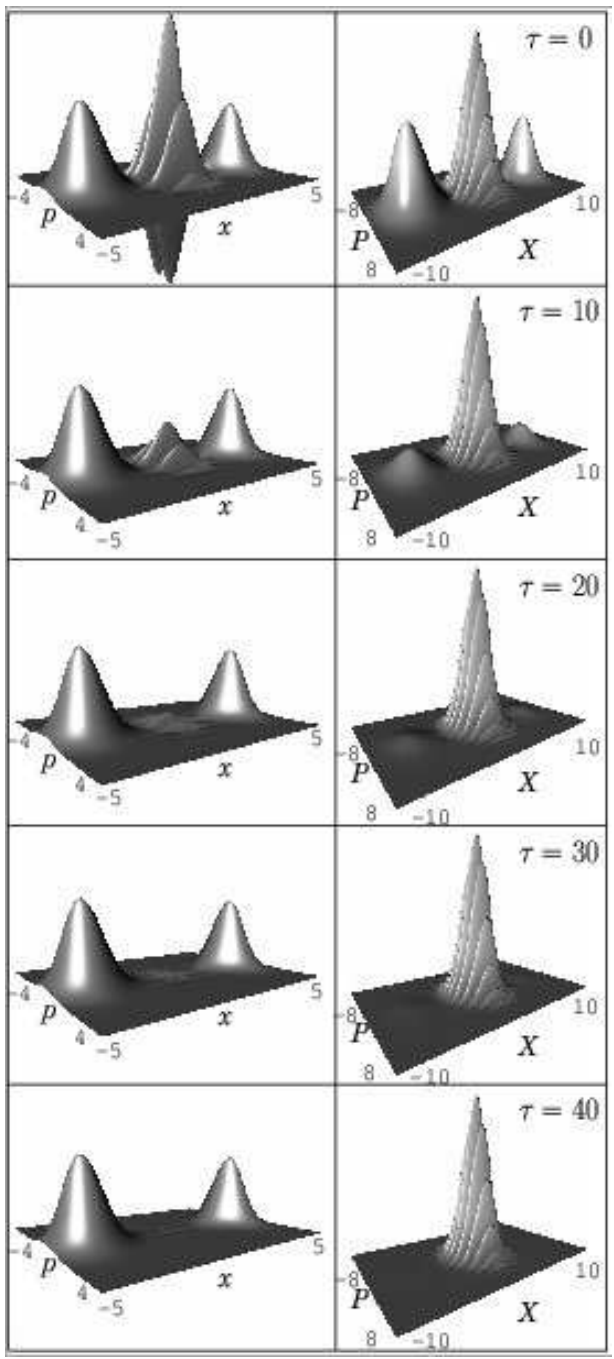}}
\end{center}
\caption{Wigner (left)  and Weyl  (right) functions  for  the ground
state of the  SQUID ring of \protect\ref{f2} evolving  in the presence
of dissipation ($T=1$K and $\protect\gamma =0.01\protect\omega $). The
effect of  dissipation is  to remove the  quantum correlations  in this superposition of macroscopically distinct states over the time shown. }
\label{f5}
\end{figure}

To  illustrate the  effect  that dissipation  has  upon a  macroscopic
superposition  state  in  a  SQUID  ring,  we  now  solve  the  master
equation~(\ref{dis}) for the system evolution, where the ring is taken
to  be in  its  ground state  at  $\Phi _{x}=0.5\Phi  _{0}$. For  this
computation  we return  to our  initial circuit  values  of $C=5\times
10^{-15}$\textrm{F}, $\Lambda  =3\times 10^{-10}$\textrm{H} and $\hbar
\nu \left( =0.047\Phi _{0}^{2}/\Lambda  \right) $. Here we have chosen
to use a decoherence rate of $0.01\omega $.  In figure~\ref{f5} we
have plotted the  Wigner and Weyl function for  the ring at particular
times.   As the  system evolves  we notice  that the  Wigner  and Weyl
functions  display  very  clearly  the disappearance  of  the  quantum
coherence between the  two states of this superposition.  We also note
that  after sufficient  time  has elapsed  the  Wigner function  still
displays two distinct flux  probabilities.  However, the SQUID ring is
longer  in a  pure state  and this  represents a  classical  coin toss
probability and not  a quantum one. This is  also clearly reflected in
the disappearance of the symmetrically positioned correlation peaks in
the Weyl function at these later times.

\subsection*{Squeezed states of a SQUID ring}

\begin{figure}[!t]
\begin{center}
\resizebox*{0.45\textwidth}{!}{\includegraphics{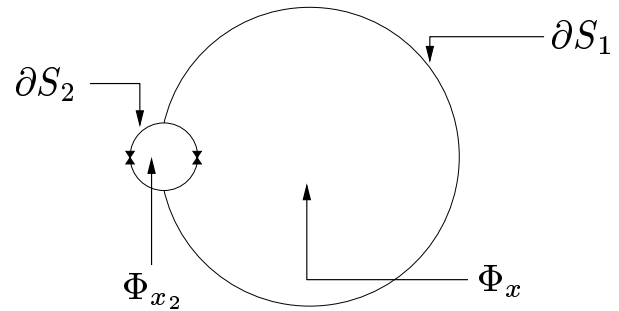}}
\end{center}
\caption{  Schematic  of  a  SQUID  ring with  a  (flux)  controllable
Josephson tunnelling energy. Here the  area enclosed by the outer loop
$\partial S_{1}$ is much greater  than that enclosed by the inner loop
$\partial S_{2}$. }
\label{fb6}
\end{figure}

Just as in quantum optics, where coherent light can be squeezed in the
number-phase plane  through its interaction with  a non-linear optical
medium, so we shall now demonstrate that a SQUID ring, with the cosine
Hamiltonian~(\ref{HamSQUID}),  can  be  used  to  squeeze  an  initial
coherent  state.  This starting  condition  in  a  SQUID ring  can  be
achieved by changing  the weak link structure in the  ring. As is well
know, the combined Josephson critical  current in a parallel, two weak
link,    loop,   connected    by   superconducting    wires    (a   DC
SQUID~\cite{Likharev86}), can be varied by adjusting the magnetic flux
threading the loop. Such a  structure can serve as the adjustable weak
link within a  larger diameter SQUID ring. This  has been already been
discussed in several earlier papers,  including that of Friedman et al
~\cite{FriedmanPCTL00}.    The     arrangement    is    depicted    in
figure~\ref{fb6}.
Provided  the control  magnetic  flux $\left(  \Phi _{x_{2}}\right)  $
threading the minor loop in  figure~\ref{fb6} is large enough, and the
weak links  in the  loop can be  fabricated (in principle)  to possess
identical critical currents, then the net supercurrent carried through
the  loop  can  be  made  vanishingly small.  In  this  situation  the
Josephson coupling  energy in (\ref{HamSQUID})  reduces essentially to
zero leaving just the parabolic  background term (i.e. as for a simple
harmonic  oscillator). Let  us now  consider a  possible  mechanism by
which we  could create a coherent state  in a SQUID ring.  If we allow
the SQUID ring to decohere to its ground state, then, as we are in the
low Josephson coupling limit, this  state will very much resemble that
of  the  vacuum state  of  the  simple  harmonic oscillator.   Now,  a
coherent state is just the  vacuum state displaced from the origin and
then allowed to evolve freely. We can realise this for a SQUID ring by
very rapidly changing the external  bias flux so that the wavefunction
does not  change significantly in shape, but  the relative expectation
value of flux  (position) no longer corresponds to  the minimum of the
potential well. We can, within reason, choose the ramp rate, start and
end bias flux, to create a state which is a very good approximation to
a coherent  state at a  required energy. We  take this as  the initial
prepared condition of the major  SQUID ring {(with a main control flux
$\Phi _{x_{1}}$)} for which we can choose a coherent state $\left\vert
A=i\sqrt{1}\right\rangle $. We follow this  {initial set up} by a very
rapid reduction  in the  {$\Phi _{x_{2}}$} to  yield the  desired (and
finite)  Josephson  coupling  energy   in  the  principal  SQUID  ring
corresponding to the ring  potential considered throughout this paper.
Clearly, in  an experimental situation this reduction  must take place
on a time scale short compared  with the decoherence time of the SQUID
ring    coupled   to    its   environment.    From    the   literature
\cite{Martinis2002,Zhou2002,Han2001,Mooij2003}  it  appears  that  the
coherence times in  SQUID rings can be sufficiently  long to make this
readily achievable. With this proviso the  system is prepared
in the  required coherent state and  then allowed to  evolve over time
using the master equation~(\ref{dis}).

\begin{figure}[!t]
\begin{center}
\resizebox*{0.45\textwidth}{!}{\includegraphics{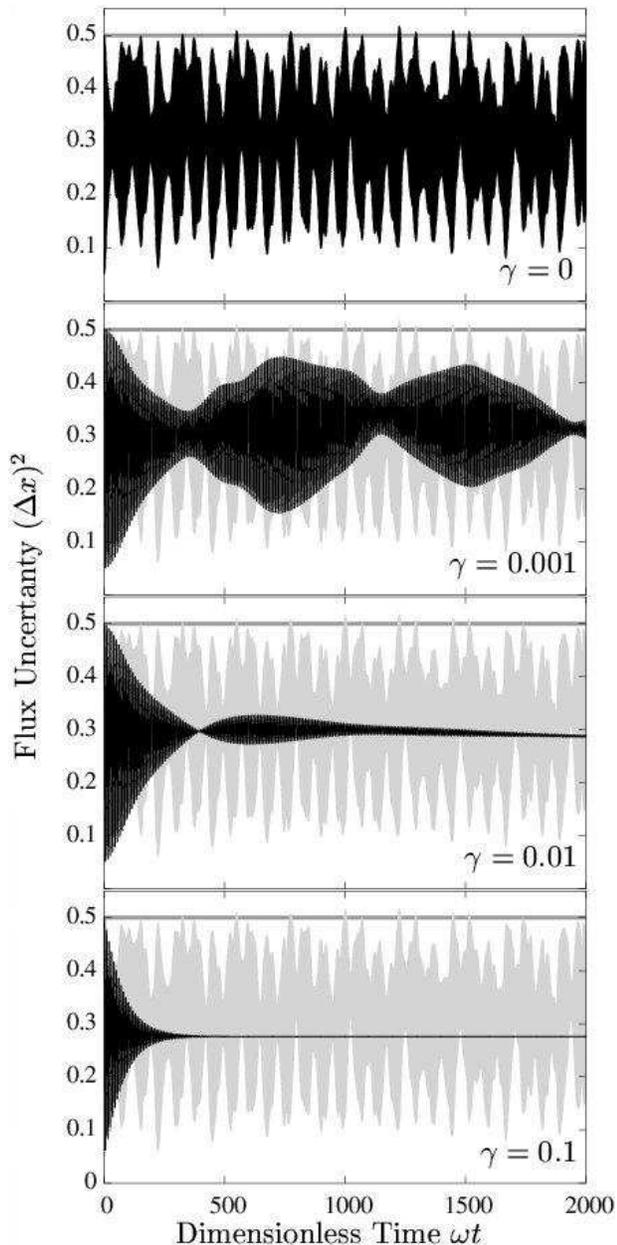}}
\end{center}
\caption{      Uncertainty     in     dimensionless      flux     (see
equation~\ref{posmom}) versus  dimensionless time ($\omega t$), for  a SQUID
ring  with parameter values  $C=5\times 10^{-15}$F,  $\Lambda =3\times
10^{-10}$H  and  $\hbar   \protect\nu  =0.24\Phi  _{0}^{2}/\Lambda  $,
computed  for a  selection of  decoherence rates  at $T=1$K.  Here the
horizontal  line  denotes  that  squeezing  occurs  for$\left(  \Delta
x\right) ^{2}$  less than $1/2$.  The grey background  oscillation for
the  $\protect\gamma=0.001,  0.01\text{  and  }0.1$  examples  is  the
$\protect\gamma=0$ pattern repeated for comparison.}
\label{f6}
\end{figure}

Here,  as before,  we take  the SQUID  ring circuit  parameters  to be
$C=5\times      10^{-15}$\textrm{F}     and      $\Lambda     =3\times
10^{-10}$\textrm{H}.  However,  in order  to  improve  the effects  of
squeezing  by the ring  we have  increased $\hbar  \nu $  to $0.24\Phi
_{0}^{2}/\Lambda  $,  yielding   a  superconducting  critical  current
density  of  around  $20\mathrm{kAcm}^{-2}$.  From our  given  initial
condition, and by computing the uncertainties in flux, we see that the
SQUID ring  can squeeze the magnetic  flux within the  ring. We choose
not to  show the  results for the  charge as  they do not  provide any
additional   information    {relevant   to   this    discussion}.   In
figure~\ref{f6}
we  show  the uncertainty  in  flux  for  a selection  of  decoherence
rates. Here,  the horizontal line  denotes that squeezing  happens for
$\left(  \Delta x\right)  ^{2}$ less  than $1/2$.  We can  clearly see
that, on  average, the  ring has effectively  squeezed this  state. We
also see that for a  judicious choice of decoherence rate and interval
of time, dissipation can assist in the squeezing process. We note that
in the $\gamma=0.1$ case it can  be seen that the SQUID ring decoheres
to its (squeezed)  vacuum state over a relatively  short period. Thus,
it appears  that these  potential wells may  be used with  facility to
generate squeezing of  the the magnetic flux variable  of the ring. It
is  therefore reasonable  to  assume that  adiabatic  changes of  this
potential  would,  unless  matching  conditions  exist  between  local
s-harmonic energy  levels of adjacent wells, allow  some adjustment of
the expectation value of the flux in the ring.

\section*{Conclusions}

In this paper  we have explored two applications  of SQUID rings, with
strong analogies to the field of quantum optics, which may prove to be
of  great  utility  given  the  current interest  in  quantum  circuit
technologies.  In the first  we describe  the manner  in which  we can
create  macroscopic superposition states  in SQUID  rings and  the way
these  may be  manipulated through  an external  control flux.  In the
second we provide a dynamical  mechanism for inducing squeezing of the
magnetic  flux  variable in  a  SQUID  ring  starting from  an  almost
harmonic  oscillator  state.  These   take  advantage  of  the  highly
non-perturbative quantum nature of SQUID rings arising from the cosine
coupling energy  term in the  ring Hamiltonian. As we  emphasise, both
are of great  interest in the light of current  research in the fields
of      quantum       computing      and      quantum      information
processing~\cite{lo_hk_1998,OrlandoMTvLLM99,MakhlinSS99,AverinNO90}. Moreover,
the results presented in this  paper emphasise that the SQUID ring can
act as  a versatile quantum device  and, as we  have shown previously,
can be  used to create correlations  across extended, multi-component,
quantum                                                         circuit
systems~\cite{EverittSCVRPP01,EverittCSPPVR01,Migliore2003a,Migliore2003b,Almaas2002,Saidi2002}. These
correlations   can,  for   example,  be   made  manifest   as  quantum
entanglements        and        quantum       frequency        up/down
conversion~\cite{EverittSCVRPP01,EverittCSPPVR01},      with      each
controlled by  the external  bias flux applied  to the SQUID  ring. In
this sense these phenomena, including macroscopic superposition states
and squeezing, highlight  the role of the SQUID  ring as the essential
machinery  for  a range  of  applications  in superconducting  quantum
circuit technologies. This further enhances our understanding of these
non-pertubative quantum objects and their possible usefulness in these
incipient technologies.

\section*{Acknowledgements}

We would like to thank the EPSRC for its generous funding of this work
and for  sponsoring the U.K.  Quantum Circuits Network. We  would also
like  to express  our thanks  both to  Professors  C.H.~van~de~Wal and
A.~Sobolev  for  interesting  discussions   and  to  the  Sussex  High
Performance  Computing  Initiative  for  the  use  of  their  NAG-IRIS
explorer graphics software.

\bibliographystyle{apsrev}
\bibliography{qcg}

\end{document}